\documentclass[aps,prl,reprint,superscriptaddress,nofootinbib,floatfix,longbibliography]{revtex4-1}
\usepackage[dvipsnames]{xcolor}
\usepackage{amssymb,amsmath,amsfonts,bm,slashed,soul,ragged2e,graphicx,epstopdf,hyperref,array}
\usepackage[utf8]{inputenc}
\usepackage{colortbl,color,caption,subcaption}
\enlargethispage{\baselineskip}
\definecolor{steelblue}{RGB}{25,25,112}
\definecolor{dullblue}{rgb}{0,0.298,0.49}
\definecolor{darkred}{rgb}{0.545,0,0}
\definecolor{darkorange}{RGB}{222,132,69}
\definecolor{darkgreen}{RGB}{126,171,85}
\definecolor{blue2}{cmyk}{1, 0.1, 0.1, 0}
\hypersetup{colorlinks,linkcolor={darkred},citecolor={blue},urlcolor={dullblue}}
\DeclareCaptionJustification{justified}{\justifying}
\everymath{\displaystyle} 

\usepackage{comment}
\usepackage{braket}
\usepackage{amsmath}
\usepackage{soul}
\usepackage{makecell}

\newcommand{\beq}{\begin{equation}}
\newcommand{\eeq}{\end{equation}}
\newcommand{\bea}{\begin{eqnarray}}
\newcommand{\eea}{\end{eqnarray}}

\newcommand{\gsim}{\lower.7ex\hbox{$\;\stackrel{\textstyle>}{\sim}\;$}}
\newcommand{\lsim}{\lower.7ex\hbox{$\;\stackrel{\textstyle<}{\sim}\;$}}

\newcommand{\be}{\begin{equation}}
\newcommand{\ee}{\end{equation}}
\newcommand{\ba}{\begin{eqnarray}}
\newcommand{\ea}{\end{eqnarray}}

\newcommand{\D}{\mathrm{d}}

\newcommand{\mi}{\mathrm{i}}
\newcommand{\me}{\mathrm{e}}

\newcommand{\lp}{\left(}
\newcommand{\rp}{\right)}

\RequirePackage[normalem]{ulem}

\begin{document}

\title{Cavity as Radio Telescope for Galactic Dark Photon}

\author{Yanjie Zeng}
\email{These two authors contribute equally.}
\affiliation{CAS Key Laboratory of Theoretical Physics, Institute of Theoretical Physics, Chinese Academy of Sciences, Beijing 100190, China}
\affiliation{School of Physical Sciences, University of Chinese Academy of Sciences, Beijing 100049, China}

\author{Yuxin Liu}
\email{These two authors contribute equally.}
\affiliation{School of Physical Sciences, University of Chinese Academy of Sciences, Beijing 100049, China}
\affiliation{International Centre for Theoretical Physics Asia-Pacific, University of Chinese Academy of Sciences (UCAS), Beijing, 100190, China}

\author{Chunlong Li}
\affiliation{School of Physics and State Key Laboratory of Nuclear Physics and Technology, Peking University, Beijing 100871, China}
\affiliation{Center for High Energy Physics, Peking University, Beijing 100871, China}

\author{Yuxiang Liu}
\affiliation{School of Physics and State Key Laboratory of Nuclear Physics and Technology, Peking University, Beijing 100871, China}
\affiliation{Center for High Energy Physics, Peking University, Beijing 100871, China}

\author{Bo Wang}
\affiliation{School of Physics, Ningxia University, Yinchuan 750021, China}

\author{Zhenxing Tang}
\affiliation{School of Physics and State Key Laboratory of Nuclear Physics and Technology, Peking University, Beijing 100871, China}
\affiliation{Beijing Laser Acceleration Innovation Center, Huairou, Beijing, 101400, China}

\author{Yuting Yang}
\affiliation{CAS Key Laboratory of Theoretical Physics, Institute of Theoretical
Physics, Chinese Academy of Sciences, Beijing 100190, China}
\affiliation{School of Physical Sciences, University of Chinese Academy of Sciences, Beijing 100049, China}
\author{Liwen Feng}
\affiliation{School of Physics and State Key Laboratory of Nuclear Physics and Technology, Peking University, Beijing 100871, China}
\affiliation{Institute of Heavy Ion Physics, Peking
University, Beijing 100871, China}

\author{Peng Sha}
\affiliation{Institute of High Energy Physics, Chinese Academy of
Sciences, Beijing 100049, China}
\affiliation{
Key Laboratory of Particle Acceleration Physics and
Technology, Chinese Academy of Sciences, Beijing 100049,
China}
\affiliation{Center for Superconducting RF and Cryogenics, Institute of
High Energy Physics, Chinese Academy of Sciences,
Beijing 100049, China}
\author{Zhenghui Mi}
\affiliation{Institute of High Energy Physics, Chinese Academy of
Sciences, Beijing 100049, China}
\affiliation{
Key Laboratory of Particle Acceleration Physics and
Technology, Chinese Academy of Sciences, Beijing 100049,
China}
\affiliation{Center for Superconducting RF and Cryogenics, Institute of
High Energy Physics, Chinese Academy of Sciences,
Beijing 100049, China}
\author{Weimin Pan}
\affiliation{Institute of High Energy Physics, Chinese Academy of
Sciences, Beijing 100049, China}
\affiliation{
Key Laboratory of Particle Acceleration Physics and
Technology, Chinese Academy of Sciences, Beijing 100049,
China}
\affiliation{Center for Superconducting RF and Cryogenics, Institute of
High Energy Physics, Chinese Academy of Sciences,
Beijing 100049, China}
\author{Tianzong Zhang}
\affiliation{School of Physics and State Key Laboratory of Nuclear Physics and Technology, Peking University, Beijing 100871, China}

\author{Zhongqing Ji}
\affiliation{Institute of Physics, Chinese Academy of Sciences, Beijing, 100190, China}

\author{Yirong Jin}
\affiliation{Beijing Academy of Quantum Information Sciences, Beijing 100193, China}

\author{Jiankui Hao}
\affiliation{School of Physics and State Key Laboratory of Nuclear Physics and Technology, Peking University, Beijing 100871, China}
\affiliation{Institute of Heavy Ion Physics, Peking
University, Beijing 100871, China}

\author{Lin Lin}
\affiliation{School of Physics and State Key Laboratory of Nuclear Physics and Technology, Peking University, Beijing 100871, China}
\affiliation{Institute of Heavy Ion Physics, Peking
University, Beijing 100871, China}
\author{Fang Wang}
\affiliation{School of Physics and State Key Laboratory of Nuclear Physics and Technology, Peking University, Beijing 100871, China}
\affiliation{Institute of Heavy Ion Physics, Peking
University, Beijing 100871, China}
\author{Huamu Xie}
\affiliation{School of Physics and State Key Laboratory of Nuclear Physics and Technology, Peking University, Beijing 100871, China}
\affiliation{Institute of Heavy Ion Physics, Peking
University, Beijing 100871, China}
\author{Senlin Huang}
\affiliation{School of Physics and State Key Laboratory of Nuclear Physics and Technology, Peking University, Beijing 100871, China}
\affiliation{Institute of Heavy Ion Physics, Peking
University, Beijing 100871, China}

\author{Yifan Chen}
\email{Corresponding author: yifan.chen@nbi.ku.dk}
\affiliation{Niels Bohr International Academy, Niels Bohr Institute, Blegdamsvej 17, 2100 Copenhagen, Denmark}

\author{Jing Shu}
\email{Corresponding author: jshu@pku.edu.cn}
\affiliation{School of Physics and State Key Laboratory of Nuclear Physics and Technology, Peking University, Beijing 100871, China}
\affiliation{Center for High Energy Physics, Peking University, Beijing 100871, China}
\affiliation{Beijing Laser Acceleration Innovation Center, Huairou, Beijing, 101400, China}

\collaboration{SHANHE Collaboration}

\begin{abstract}
Dark photons, as a minimal extension of the Standard Model through an additional Abelian gauge group, may propagate relativistically across the galaxy, originating from dark matter decay or annihilation, thereby contributing to a galactic dark photon background. The generation of dark photons typically favors certain polarization modes, which are dependent on the interactions between dark matter and dark photons. We introduce a framework in which a resonant cavity is utilized to detect and differentiate these polarizations, leveraging the daily variation in expected signals due to the anisotropic distribution of dark photons and the rotation of the Earth. We conduct an experimental search using superconducting radio-frequency cavities, noted for their exceptionally high quality factors, proving them to be effective telescopes for observing galactic dark photons. This approach establishes the most stringent limits yet on the kinetic mixing coefficient between dark photons and electromagnetic photons, thereby unveiling a novel avenue for the indirect search for dark matter via multi-messenger astronomy.
\end{abstract}

\date{\today}

\maketitle

\noindent{\bf 1. Introduction\;}

Dark photon, a new vector field posited as a minimal extension to the Standard Model (SM), arises from speculative new physics theories such as extra dimensions or grand unification~\cite{Svrcek:2006yi,Abel:2008ai,Arvanitaki:2009fg,Goodsell:2009xc}. When present in large occupation numbers, dark photons can form a wave-like background, inducing an effective electric current through mixing with electromagnetic photons~\cite{Horns:2012jf,Chaudhuri:2014dla}. To probe this intricate interaction, various experimental approaches have been undertaken. Notably, superconducting radio-frequency (SRF) cavities have shown exceptional promise, offering the most stringent constraints thus far~\cite{Cervantes:2022gtv,Romanenko:2023irv,SHANHE:2023kxz} due to their extremely high quality factor.

The dark photon can itself serve as a dark matter candidate, characterized by its wave-like nature at sub-eV mass scales~\cite{Nelson:2011sf, Arias:2012az}. Beyond constituting the majority of cold dark matter in a standard halo distribution, dark photons may also arise in diverse cosmic scenarios~\cite{Chen:2021bdr}, such as forming an isotropic cosmological background.
Anisotropic dark photons can originate from galactic sources including supernovae~\cite{Raffelt:1999tx,Eby:2024mhd,deGiorgi:2024pjb}, bursts from boson clumps~\cite{Eby:2016cnq, Levkov:2016rkk, Budker:2023sex,Du:2023jxh}, or ionization processes~\cite{Baumann:2021fkf,Duque:2023cac}, and dipole radiation emitted by binary systems~\cite{Cardoso:2016olt, Dror:2021wrl}. Notably, dark photons produced via dark matter decay or annihilation contribute to a galactic background with a distinct directional flow from the galactic center. 
The detection of these galactic dark photons thus serves as an indirect detection of dark matter. Their anisotropic distribution on Earth results in diurnal signal modulation in direction-sensitive detectors~\cite{ADMX:2023rsk}, which can be utilized as distinctive indicators of their galactic origins. Moreover, analyzing the polarization and spectral characteristics of dark photons offers in-depth insights into the nature of dark matter interactions.

In this work, we initiate the search for galactic dark photons using an SRF cavity. We demonstrate that projecting dark photon wavefunctions onto the cavity's modes results in polarization-dependent modulation of signatures, effectively transforming the cavity into a radio-telescope for dark photons. Utilizing a tunable SRF cavity, cooled in a liquid helium environment to approximately $2\,$K and connected to several amplifiers, we establish pioneering constraints on galactic dark photons across two distinct polarization modes. Our analytical method is broadly applicable to anisotropic dark photon fluxes, focusing specifically on two benchmark models of dark matter decay: a flat spectrum from four-body cascade decay~\cite{ADMX:2023rsk} and a narrow spectrum, $1/500$ of the central frequency, associated with two-body parametric decay. We explore a vast, previously untapped parameter space, leveraging the exceptional quality factor of the SRF cavity, which provides unique advantages over traditional copper cavities~\cite{ADMX:2023rsk}.

\vskip 0.7cm

\noindent{\bf 2. Galactic Dark Photon Flux \;}

We begin by characterizing the wavefunctions and correlations inherent to a general dark photon background. The wavefunction $X^\mu$ is formulated in the frequency domain as:
\begin{equation}
    X^{ \mu} (t,\bm{x}) = \int \sum_P\, X_{P}(\omega,\hat{\bm{k}}) \,e^{\mu}_{P}\,\me^{\mi(\omega t- \bm{k}\cdot \bm{x})}\,\frac{\D \omega}{2\pi} \, \D^2 \hat{\bm{k}},
\end{equation}
where the frequency $\omega$ and spatial momentum $\bm{k}$ adhere to the dispersion relation $\omega^2=|\bm{k}|^2+ m_{X}^2$ for the dark photon mass $m_X$. $X_P$ signifies the amplitude for each polarization mode $P$, incorporating one longitudinal (L) and two transverse (T) components, while $e^{\mu}_{P}$ is the associated polarization vector basis.

Considering the non-coherent nature of dark matter decay, galactic dark photons are expected to manifest as a stochastic Gaussian field. This is quantified by the two-point correlation function:
\begin{equation}
\langle X_P(\omega,\hat{\bm{k}}) X_{P'}(\omega',\hat{\bm{k}}')^* \rangle = \delta(\omega-\omega') \delta^2(\hat{\bm{k}}-\hat{\bm{k}}')\delta_{PP'} \mathcal{S}_P(\omega,\hat{\bm{k}}).
\end{equation}
Here, $\langle \cdots \rangle$ denotes the ensemble average and $\mathcal{S}_P(\omega,\hat{\bm{k}})$ is the power spectral density (PSD) of the $P$-mode dark photon from direction $\hat{\bm{k}}$. The total energy density of the dark photon is obtained by integrating $\omega^2 \mathcal{S}_P(\omega,\hat{\bm{k}})$ over the frequency spectrum.

The polarization states and angular distribution of the dark photon spectrum depend on its production mechanisms, as detailed in the Supplemental Material. We examine scenarios where either longitudinal or transverse modes are predominant. Specifically, the longitudinal mode often arises in dark Higgs models, where a scalar field $\Phi$ decays into two relativistic dark photons through an interaction Lagrangian proportional to $\Phi X^{\mu}X_{\mu}$. Conversely, transverse modes are typically generated through axion-photon-like interactions, denoted by a term proportional to $\Phi X_{\mu \nu} \tilde{X}^{\mu \nu}$, where $X_{\mu \nu}$ represents the dark photon field tensor and $\tilde{X}^{\mu \nu}$ its dual.

\begin{figure}[t]
    \centering
       \includegraphics[width=0.45\textwidth]{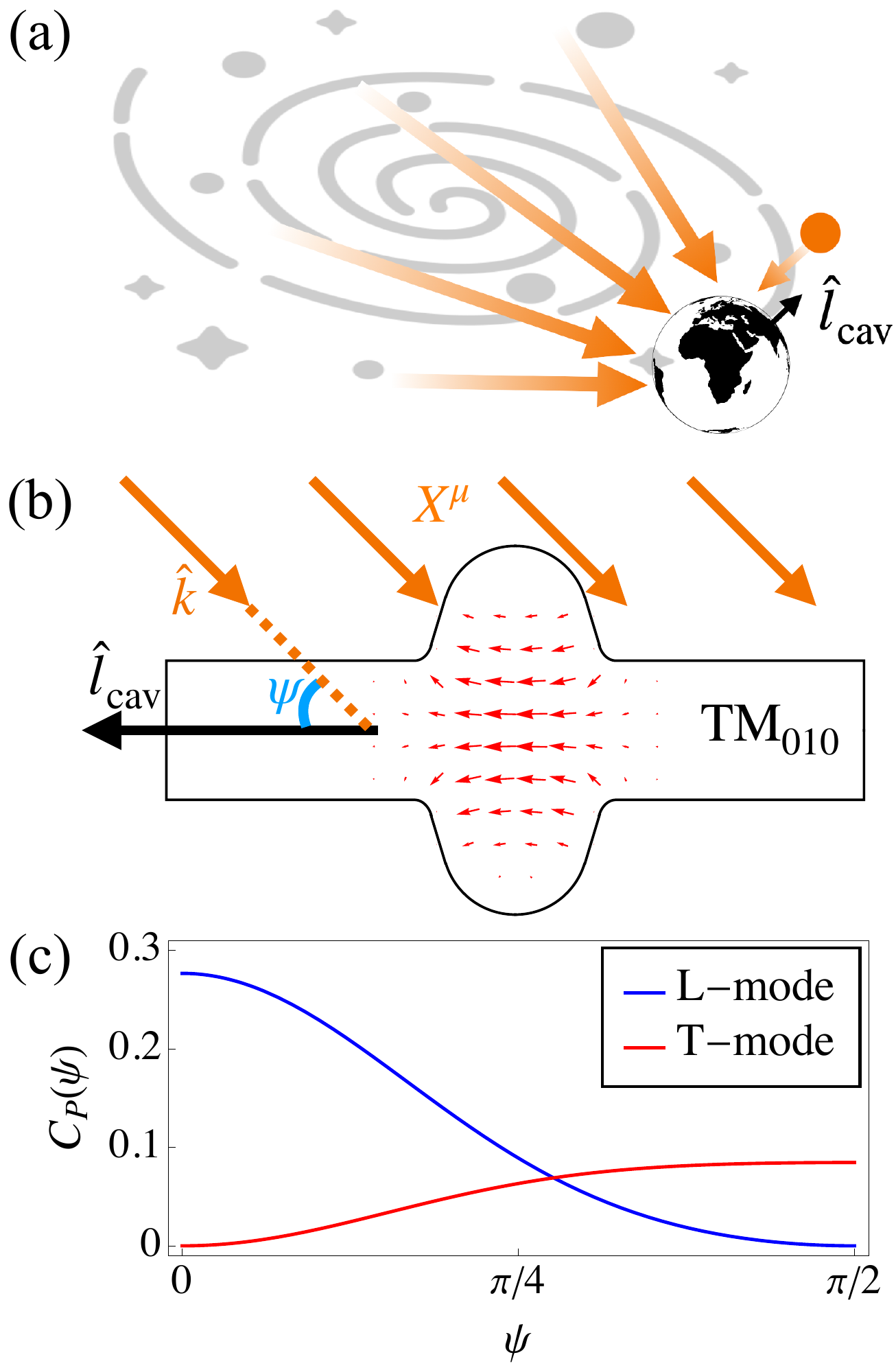}
    \caption{{\bf (a):} Illustration of anisotropic dark photon fluxes propagating from galactic sources, depicted by orange arrows, and a nearby source, indicated by an orange circle. The symbol $\hat{\bm{l}}_{\rm cav}$ denotes the symmetry axis of the cavity, which is oriented towards the Earth’s center.
    {\bf (b):} Schematic of the single-cell $1.3$\,GHz elliptical cavity utilized in this study with incoming dark photon fields from direction $\hat{\bm{k}}$ (orange). The angle between $\hat{\bm{k}}$ and the cavity's symmetry axis $\hat{\bm{l}}_{\rm cav}$ is denoted as $\Psi$. The electric fields of the targeted cavity mode TM$_{010}$ are illustrated with red arrows, their length proportional to the field strength. 
    {\bf (c):} The overlap factor between the TM$_{010}$ mode and incoming dark photon fields of different polarization modes, as defined in Eq.\,(\ref{eq:C}), assuming $|\bm{k}|=\omega_0$ for a relativistic dark photon.
    }
    \label{fig:C}
\end{figure}

We consider two benchmark scenarios in which dark matter decays produce such fluxes, corresponding to two extreme source bandwidths relevant for resonant cavity detection.
For the interaction models under consideration, the decay of a massive scalar field into two dark photons can induce parametric instability, leading to rapid depletion of the scalar field through Bose-enhanced dark photon production~\cite{Dror:2021nyr}. In conventional virialized scalar halos, such instabilities might be moderated by gravitational redshift when couplings are sufficiently weak~\cite{Arza:2020eik}, albeit at the cost of reducing the decay products to near indetectable levels. To address this, we consider that a fraction of the scalar dark matter aggregates into dense clumps nested within broader, more diffuse miniclusters~\cite{hogan1988axion,Kolb:1993hw}. Within these dense regions, an equilibrium between dark photon production and escape, limited by the clumps' size, culminates in a saturated state of dark photons. The clumps achieve stability via a balance of parametric decay and accretion, resulting in a steady and substantial dark photon flux that eclipses that from mere perturbative decay, akin to saturated superradiant clouds in axion-photon systems~\cite{Rosa:2017ury,Spieksma:2023vwl}. The parametric instability band predominantly shapes the resulting dark photon spectrum, related to the scalar clumps' core size~\cite{Hertzberg:2018zte,Levkov:2020txo}. Assuming a virialized velocity distribution with a dispersion of $10^{-3}\,c$, the produced dark photons exhibit a bandwidth approximately $1/500$ of the central frequency. An alternative production mechanism involves a scalar dark matter field decaying into two intermediate states, $\Phi$, which then undergo cascade decay into four dark photons~\cite{ADMX:2023rsk}. Here, the dark photon spectrum is broadly flat, confined within a frequency range dictated by the dark matter mass.

In both scenarios, local dark photon flux densities may soar to $1000\,\rho_\gamma$, compatible with cosmological constraints on dark matter decay~\cite{Dror:2021nyr,ADMX:2023rsk}, where $\rho_\gamma$ signifies the local electromagnetic photon density~\cite{Workman:2022ynf}. 
The angular distribution of the incoming dark photon flux corresponds to the line-of-sight integral of the dark matter density, as depicted in Fig.\,\ref{fig:C}a. This parameter is commonly referred to as the J-factor in the literature on indirect dark matter detection\cite{Cirelli:2010xx}. It is important to note that certain nearby sources, indicated by the orange circle in Fig.\,\ref{fig:C}a, may produce significantly stronger transient fluxes.

\vskip 0.7cm

\noindent{\bf 3. Cavity as a Dark Photon Telescope\;}

When dark photon fluxes transit through a cavity, they can potentially excite a cavity mode if their frequency aligns with the resonant frequency of the cavity, a process facilitated by kinetic mixing with electromagnetic photons~\cite{SHANHE:2023kxz}. In the interaction picture, dark photons contribute to an effective electric current $\bm{J}_{\text{eff}} = \epsilon\, m_{X}^2 \bm{X}$, where $\epsilon$ denotes the kinetic mixing coefficient, and $\bm{X}$ represents the spatial component of the dark photon wavefunction. As elucidated in Supplemental Material, the rate of excitation depends on the overlap between $\bm{X}$ and the cavity mode's electric field $\bm{E}_0$, quantified by the parameter:
\begin{equation}
    C_{P}(\hat{\bm{k}}) \equiv \frac{1}{V} \left\vert\int_V  \me^{-\mi \bm{k}\cdot \bm{x}} \, \hat{\bm{e}}_{P}(\hat{\bm{k}})\cdot\bm{E}_0(\bm{x}\,) \,\D^3 \bm{x} \,  \right\vert^2,
    \label{eq:C}
\end{equation}
where $\hat{\bm{e}}_{P}$ is the spatial unit directional vector of the polarization basis $e_P^\mu$, and the integral is taken over the cavity's volume $V$. Typically, the phase factor $\me^{-\mi \bm{k}\cdot \bm{x}}$ is approximated as $1$ for the overlap function for DPDM~\cite{SHANHE:2023kxz}, due to the de-Broglie wavelength $1/|\bm{k}|$ of the non-relativistic field being considerably longer than the cavity scale. For axisymmetric cavity modes, the angular dependence reduces to the polar angle $\Psi \equiv -\hat{\bm{k}}\cdot\hat{\bm{l}}_{\rm cav}$, with $\hat{\bm{l}}_{\rm cav}$ representing the cavity's symmetry axis, as depicted in Fig.\,\ref{fig:C}b. We numerically evaluate the overlap factor $C_{P}(\Psi)$ for the TM$_{010}$ mode of a single-cell $1.3$\,GHz elliptical cavity utilized in this study, illustrated in Fig.\,\ref{fig:C}c. The longitudinal mode, with $\hat{\bm{e}}_{L}(\hat{\bm{k}})$ aligned with $\hat{\bm{k}}$, achieves maximum overlap at $\Psi=0$ and diminishes to zero at $\Psi=\pi/2$. The two transverse circular polarized modes, indistinct from one another, display an inverse variation compared to the longitudinal mode. These evaluations assume a relativistic limit for the dark photon, where $|\bm{k}|\approx \omega_0$.

The PSD absorbed by the cavity from dark photon fields bears resemblance to that from DPDM~\cite{SHANHE:2023kxz}:
\begin{equation}
    S_{\mathrm{sig}}^P(\omega,t) = \frac{\epsilon^2\, m_{X}^4\, V\, |\bm{e}_P|^2}{4(\omega-\omega_0)^2 + \omega_0^2/Q_L^2}
    \int  \mathcal{S}_P(\omega,\hat{\bm{k}}) C_{P}(\Psi(t)) \D^2 \hat{\bm{k}}.
\label{eq:sig_PSD}
\end{equation}
The distinction comes with the factor $|\bm{e}_P|^2$ denotes the magnitude square of the spatial component of the polarization basis $e_P^\mu$. This factor equates to $1$ for transverse modes and escalates to $\omega^2/m_X^2$ for the longitudinal mode, suggesting that a relativistic longitudinal mode could significantly enhance signal power compared to transverse modes, given the same spectra. Here, $Q_L$ indicates the cavity's loaded quality factor, determining the resonance bandwidth. In the Galactic frame, the Earth's rotation leads to temporal variations in the cavity's symmetry axis $\hat{\bm{l}}_{\rm cav}(t)$ and the associated $\Psi(t)$. Consequently, these changes result in polarization-dependent modulations in the PSD, which can be leveraged as distinct signatures for the detection of anisotropic dark photons.

\vskip 0.7cm

\noindent{\bf 4. Diurnal Modulation Analysis\;}

We now embark on a search for the diurnal modulation of signals stemming from galactic dark photons, leveraging a single-cell elliptical niobium SRF cavity~\cite{SHANHE:2023kxz}. The cavity was equipped with a frequency tuning mechanism~\cite{Pischalnikov:2015eye,Pischalnikov:2019iyu} and an amplifier circuit, and maintained in liquid helium at a temperature of approximately $2$\,K. The experimental platform~\cite{Mi:2014wwa,Liu:2016lzs,Zhou:2018vcm,Zhang:2022bmh,YANG20221354092,app13158618} facilitated the calibration of relevant parameters, including the loaded quality factor $Q_L$, coupling factor $\beta$, net amplification factor $G_{\rm net}$, and the resonant frequency of the TM$_{010}$ mode, centered around $\omega_0/(2\pi)\approx  1.3$\,GHz, along with its stability range for each scanning process. Each scan step was approximately $5$\,minutes, with $100$\,seconds dedicated to data acquisition and the remainder to frequency tuning and parameter calibration. During two continuous operational periods from April 12-14 and 23-26, 2023, we completed $N_{\rm bin} = 350$ and $800$ scan steps respectively, covering a total frequency range of $1.37$\,MHz.

\begin{figure}[t]
    \centering
    \includegraphics[width=0.48\textwidth]{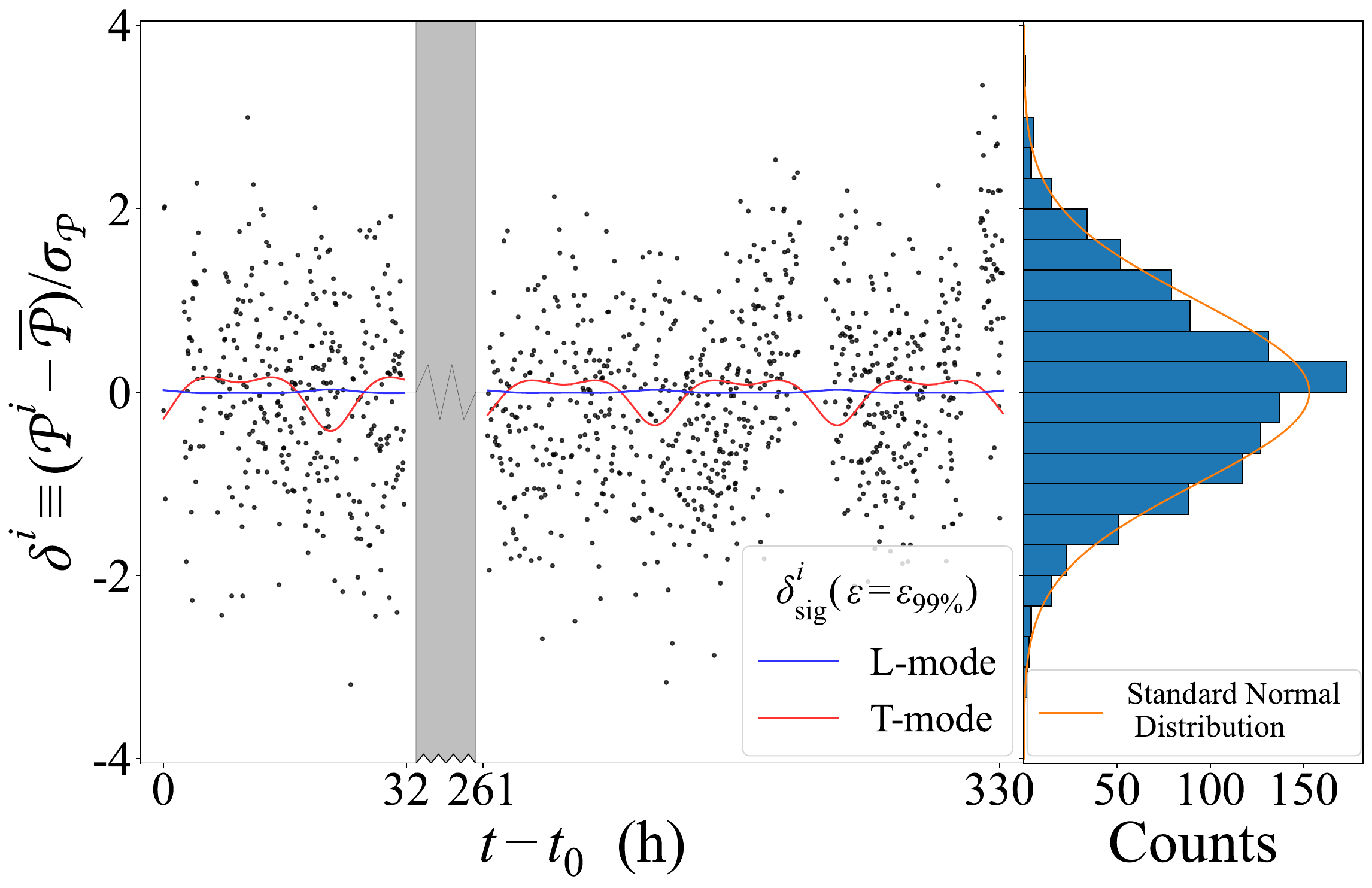}
    \caption{Normalized power excess from experimental data (black dots) for each scan step as a function of measurement time, and the predicted modulation signals from galactic dark photons for two polarization modes (blue and red lines, respectively). The right panel shows the distribution of normalized power excess, which fits a Gaussian distribution with a mean of $< 10^{-4}$, a standard deviation of $0.9991$, and a p-value of $0.80$, with no $4\sigma$ excess observed.
    The experiment's initial time, $t_0$, was UTC 22:23:52, April 12, 2023. The two runs are delineated by the gray region. The modulation signals are customized for the detector's location ($116.4^{\circ}\mathrm{E}$, $39.9^{\circ}\mathrm{N}$, Beijing) and orientation (cavity's symmetry axis pointing towards Earth's center), with the signal normalization set to the $90\%$ exclusion bound.
    }
    \label{fig:mod}
\end{figure}

In our analysis, only the received power at the resonant frequency bin with a bandwidth of $\Delta \omega_0 /(2\pi)= 11.5$\,Hz~\cite{SHANHE:2023kxz}, denoted by $\mathcal{P}^i$ for the $i$-th step, was considered. The dominant noise contribution typically comes from the amplifier. For each run, we computed the average and standard deviation of $\mathcal{P}^i$, denoted as $\overline{\mathcal{P}}$ and $\sigma_\mathcal{P}$, respectively, and determined the normalized power excess $\delta^i \equiv (\mathcal{P}^{i}-\overline{\mathcal{P}})/{\sigma_\mathcal{P}}$. These are shown as black dots in Fig.\,\ref{fig:mod}, aligning closely with a standard normal distribution with a mean of $< 10^{-4}$, a standard deviation of $0.9991$, and a p-value of $0.80$, with no $4\sigma$ excess observed.

Unlike in the scan search for dark photon dark matter, where a constant fit and normalized power excess were calculated across every $50$ contiguous bins to ensure environmental stability~\cite{SHANHE:2023kxz}, the current search combines all bins from both runs without deliberately suppressing potential diurnal variations.

The expected power signal induced by galactic dark photons at scan step $i$ is given by:
\be     \mathcal{P}^{i}_{\rm sig}(t) = G_{\rm net}\,\eta_{\rm bin} \frac{\beta}{\beta+1}\frac{(\omega_0^i)^2}{8 \pi \, Q_L^2} \sum_P S_{\rm sig}^P(\omega_0^i,t).\ee
Here, $\omega_0^i$ is the cavity's resonant frequency, and $\eta_{\rm bin} \simeq 84\%$ accounts for the signal leakage outside the bin due to variations in the resonant frequency. As the total scan range of $ 1.37$\,MHz is less than the considered galactic dark photons' bandwidth, we assume the spectrum $\mathcal{S}_P(\omega_0^i,\hat{\bm{k}})$ in Eq.\,(\ref{eq:sig_PSD}) remains constant with respect to $\omega_0^i$. The diurnal signal variation is calculated by considering the detector's location ($116.4^{\circ}\mathrm{E}$, $39.9^{\circ}\mathrm{N}$, Beijing) and orientation (symmetry axis pointing towards Earth's center), along with Earth's rotational changes in the Galactic frame, as detailed in Supplemental Material. The time variation during each scan is negligible compared to the diurnal variation timescale. Analogous to $\delta^i$, we define $\delta_{\rm sig}^i \equiv (\mathcal{P}^{i}_{\rm sig}-\Sigma_i \mathcal{P}^{i}_{\rm sig}/N_{\rm bin})/\sigma_\mathcal{P}$ for the signal variations, exemplified in Fig.\,\ref{fig:mod} for the two polarization modes.

\begin{figure}[t]
    \centering
    \includegraphics[width=0.48\textwidth]{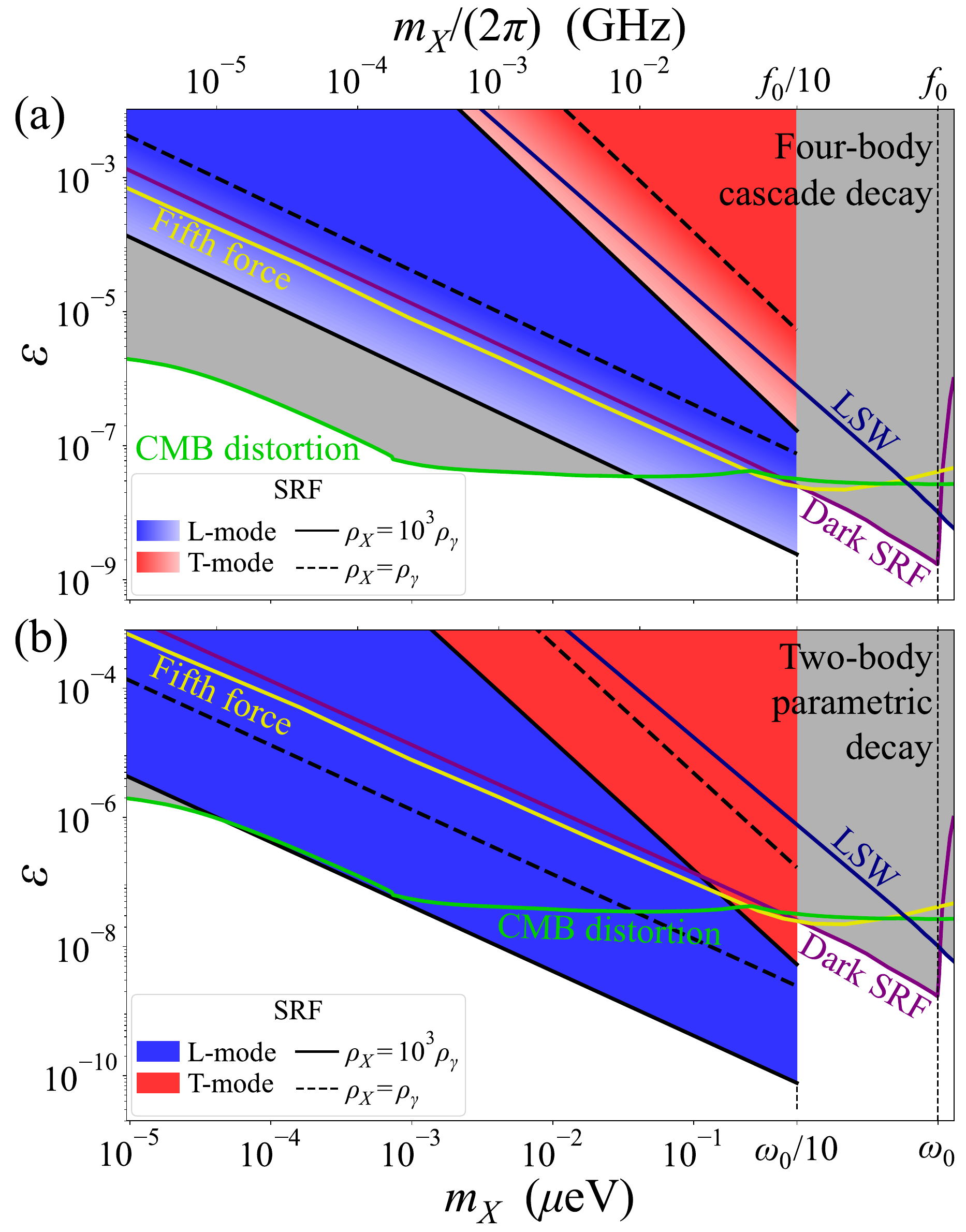}
    \caption{The $90\%$ exclusion limits on the kinetic mixing coefficient $\epsilon$ for galactic dark photons with mass $m_X$. The sub-figures correspond to the two production mechanisms: four-body cascade decay (a) and two-body parametric decay (b), respectively. Exclusions for the longitudinal and transverse polarization modes are shown in blue and red. 
    The filled region assumes $\rho_X = 1000\,\rho_\gamma$, while the black dashed lines correspond to $\rho_X = \rho_\gamma$, with the exclusion boundary scaling as $1/\sqrt{\rho_X}$.
    For context, we include established constraints from light-shining-through-walls (LSW) experiments like Dark SRF~\cite{Romanenko:2023irv} using similar SRF cavities, and earlier results~\cite{Betz:2013dza}, fifth-force searches~\cite{Kroff:2020zhp}, and CMB distortions from photon-dark photon oscillations~\cite{McDermott:2019lch,Caputo:2020bdy}. The gradient color in the four-body cascade decay panel indicates variations in the dark matter particle mass from $4\,\omega_0$ (light) to $100\,\omega_0$ (dark). For the two-body parametric decay, the dark matter mass is set at approximately $2\,\omega_0$. }
    \label{fig:90exclusion}
\end{figure}

We evaluate the hypothesis of galactic dark photons characterized by mass $m_X$ and kinetic mixing coefficient $\epsilon$ utilizing the following probability function:
\begin{equation}
    \mathrm{Pr}\left(\delta^i \mid \epsilon, m_{X}\right) \propto \prod_{i} \exp \left( \frac{-\left( \delta^i -\delta_{\rm sig}^i \right)^2}{2 +2(\delta^i)^2\sum_j F_j^{2}} \right),
\label{eq:PR}
\end{equation}
where $F_j^2$ quantifies the uncertainties of the involved parameters, collectively accounting for less than $5\%$~\cite{SHANHE:2023kxz}, as detailed in Supplemental Material. The $90\%$ exclusion limit derived from this probability function is depicted in Fig.\,\ref{fig:90exclusion}, showcasing constraints for the two benchmark spectra: a flat spectrum from four-body cascade decay, and a $1/500$ bandwidth around the central frequency of approximately $\omega_0$ from two-body parametric decay. 
Both spectra are normalized to a dark photon energy density of $\rho_X = 1000\, \rho_\gamma$. 
We constrain our analysis to regions where dark photons are relativistic, specifically where $m_X\leq \omega/10$, as non-relativistic regions lead to a mix of polarization modes and diminish the diurnal modulation. Notably, the longitudinal mode is more significantly constrained than the transverse modes, attributed to the distinct factors $|\bm{e}_L|^2 = \omega^2/m_X^2$ and $|\bm{e}_T|^2 = 1$ in the signal power expression (Eq.\,(\ref{eq:sig_PSD})). The constraints for the four-body cascade decay also depend on the mass of the original dark matter particle, which influences the bandwidth of $\mathcal{S}_P$ and thus the amplitude at the cavity's resonant frequency. The gradient color region in the figure indicates the variation in exclusion for dark matter masses ranging from $4\,\omega_0$ (light) to $100\,\omega_0$ (dark). Alternatively, the two-body parametric decay leads to dark photons with energy half that of the dark matter particle mass, requiring the latter to be around $2\,\omega_0$ for our analysis. The dark photon density may be lower than the assumed $\rho_X = 1000\, \rho_\gamma$. For comparison, we include the case $\rho_X = \rho_\gamma$, shown as black dashed lines, with the exclusion boundary scaling as $1/\sqrt{\rho_X}$.

We also juxtapose our findings with established constraints on kinetic-mixing dark photons from various sources. This includes light-shining-through-walls (LSW) experiments such as Dark SRF~\cite{Romanenko:2023irv}, alongside a spectrum of previous results~\cite{Betz:2013dza, Parker:2013fxa}, combined constraints from fifth-force searches~\cite{Kroff:2020zhp}, and the effects on the cosmic microwave background (CMB) from photon-dark photon oscillations~\cite{McDermott:2019lch,Caputo:2020bdy}.

To better understand the scaling of the exclusion depicted in Fig.\,\ref{fig:90exclusion}, we employ an approximation based on the Dicke radiometer equation~\cite{Dicke:1946glx}: ${\rm SNR}^2= \Sigma_i (\delta^i_{\rm sig})^2 t_{\rm int}\Delta\omega_0/(2\pi)$. Setting a requirement for the SNR$^2$ to reach $5$ leads to the following estimation:
\begin{equation}
\label{eq:epsilon}
\begin{aligned}
    \epsilon \approx \, &4\times10^{-8}\left(\frac{0.1\mu{\rm eV}}{m_X}\right)^2\left(\frac{1000\,\rho_\gamma}{\rho_X}\right)^{\frac{1}{2}}\left(\frac{500}{Q_X}\right)^{\frac{1}{2}}\\
    &\left(\frac{0.1}{\bar{C}_P}\right)^{\frac{1}{2}}\left(\frac{10\%}{\mathcal{F}_{C_P}}\right)^{\frac{1}{2}}\left(\frac{T_{N}}{3\, \rm{K}}\right)^{\frac{1}{2}}\left(\frac{1}{|\bm{e}_P|^2}\right)^{\frac{1}{2}}\left(\frac{100\,{\rm s}}{t_{\rm int}}\right)^{\frac{1}{4}}\\
    &\left(\frac{1000}{N_{\rm bin}}\right)^{\frac{1}{4}}\left(\frac{4\,{\rm L}}{V}\right)^{\frac{1}{2}}\left(\frac{\omega_0/(2\pi)}{1.3\,\rm{GHz}}\right)^{\frac{3}{2}}\left(\frac{\Delta \omega_0/(2\pi)}{10\,{\rm Hz}}\right)^{\frac{1}{4}},
\end{aligned}
\end{equation}
where $\bar{C}_P$ denotes the average overlapping factor during modulation, $\mathcal{F}_{C_P}$ represents the modulation factor of $\delta_{\rm sig}^i$, defined as the ratio of the standard deviation to the average, and $T_N$ indicates the noise temperature of the amplifier. All parameters approximately follow the experiments and models considered. Equation~(\ref{eq:epsilon}) elucidates that significant sensitivity improvements are achievable through the narrower bandwidth of the SRF cavity, attributable to its exceptionally high quality factor.

\vskip 0.7cm

\noindent{\bf 5. Conclusion\;---}
In this study, we have investigated galactic dark photons as decay products from scalar dark matter, demonstrating the SRF cavity's sensitivity to their largely unexplored parameter space in terms of mass and kinetic mixing coefficient.
The two production mechanisms we analyzed allow for ultralight bosonic dark matter decay without exponential depletion of the background, thus providing a novel avenue for the indirect detection of scalar dark matter via dark photons. Due to their anisotropic distribution, the induced signal in a cavity at rest relative to the Earth exhibits diurnal modulation, a feature previously used to search for Earth-scattering effects from heavier, boosted dark matter~\cite{Ge:2020yuf,Fornal:2020npv,Chen:2021ifo,Xia:2021vbz,PandaX-II:2021kai,Qiao:2023pbw}. This modulation pattern is sensitive to the polarization mode of dark photons, with constraints being particularly stringent for the longitudinal polarization mode. 
The exceptional sensitivity of our study is attributed to the extremely high quality factor of the SRF cavity. This high quality factor results in a narrow bandwidth, which inherently supports low noise levels, thus enabling the exclusion of galactic dark photons across a spectrum bandwidth substantially broader than the cavity's own bandwidth. This approach mirrors the methodology employed in DPDM searches~\cite{Cervantes:2022gtv,Romanenko:2023irv,SHANHE:2023kxz}, showcasing the effectiveness of SRF cavities in probing wide spectral ranges of dark photon signatures.

Looking ahead, as more cavities become operational, a network of detectors with extensive long-baseline separation~\cite{Foster:2020fln, Chen:2021bdr, GNOME:2023rpz, Jiang:2023jhl} will enable detailed analysis and precise localization of dark photons, including their polarization modes and chiral components~\cite{Chen:2021bdr}. This approach not only enhances our understanding of dark photon properties at both macroscopic and microscopic levels but also, when integrated with other astrophysical observations such as gravitational waves, opens new frontiers in multi-messenger astronomy and cosmology~\cite{Dailey:2020sxa}. Beyond dark photon detection, the versatility of SRF cavities extends to axion searches and high-frequency gravitational wave detection through heterodyne upconversion~\cite{Goryachev:2018vjt, Berlin:2019ahk, ADMX:2023rsk}. Moreover, enhancements in SRF cavity sensitivity can be achieved through advanced readout systems~\cite{Dixit:2020ymh, Agrawal:2023umy, HAYSTAC:2020kwv, Li:2020cwh, Chen:2021bgy, Wurtz:2021cnm, Jiang:2022vpm, Chen:2023ryb}.

\bigskip

\noindent{\bf Conflict of interest}
The authors declare that they have no conflict of interest.
\hspace{5mm}
\begin{acknowledgements}
\noindent{\bf Acknowledgements}
We are grateful to Nicholas L. Rodd and Hyeonseok Seong for useful discussions. We acknowledge the utilization of the Platform of Advanced Photon Source Technology R\&D. This work is supported by the National Key Research and Development Program of China under Grant No. 2020YFC2201501.
Yifan Chen is supported by VILLUM FONDEN (grant no. 37766), by the Danish Research Foundation, and under the European Union’s H2020 ERC Advanced Grant “Black holes: gravitational engines of discovery” grant agreement no. Gravitas–101052587, and by the European Consortium for Astroparticle Theory in the form of an Exchange Travel Grant, and by FCT (Fundação para a Ciência e Tecnologia I.P, Portugal) under project No. 2022.01324.PTDC.
Jing Shu is supported by the National Natural Science Foundation of China under Grants No. 12025507,  No.12450006, No.12150015 and Peking University under startup Grant No. 7101302974; and is supported from the State Key Laboratory of Nuclear Physics and Technology, Peking University (No. NPT2024ZX01).
\end{acknowledgements}

\noindent{\bf Author Contributionjias}
Jing Shu established and coordinated the collaboration. Yifan Chen and Jing Shu conceived the study. Yanjie Zeng and Chunlong Li participated in the data analysis, with help from Yifan Chen and Bo Wang. Yuxin Liu developed dark photon production mechanisms. Yanjie Zeng and Yuxiang Liu calculated the cavity response to dark photon production. Yifan Chen and Jing Shu coordinated the draft writing, with contributions from Yuxin Liu and Yanjie Zeng. Bo Wang, Zhenxing Tang, Yuting Yang, Liwen Feng, Yanjie Zeng, Chunlong Li, Tianzong Zhang, Peng Sha, and Zhenghui Mi participated in the experimental searches, with help from the whole collaboration, including Weiming Pan, Zhongqing Ji, Yirong Jin, Jiankui Hao, Lin Lin, Fang Wang, Huamu Xie, and Senlin Huang. All authors contributed to the discussion.

%

\appendix
\pagebreak
\widetext
\begin{center}
\textbf{\large Supplemental Materials: Cavity as Radio Telescope for Galactic Dark Photon}
\end{center}
\setcounter{equation}{0}
\setcounter{figure}{0}
\setcounter{table}{0}
\makeatletter
\renewcommand{\theequation}{S\arabic{equation}}
\renewcommand{\thefigure}{S\arabic{figure}}
\renewcommand{\thetable}{S\arabic{table}}
\renewcommand{\bibnumfmt}[1]{[#1]}
\renewcommand{\citenumfont}[1]{#1}



\section*{I: Cavity Response to General Dark Photon Background}

This section delves into a comprehensive calculation of the signal generated in a cavity by a general background of kinetically mixing dark photons. The background is characterized by an angular-dependent power spectral density (PSD) function. Unlike non-relativistic dark photon dark matter (DPDM), 
a relativistic dark photon can exhibit a position-dependent phase factor in the cavity signal, owing to its de Broglie wavelength being comparable to the cavity size.

\subsection*{A. General Dark Photon Background}

We examine a dark photon $X^\mu$ with mass $m_X$ and momentum $\bm{k}$ in the Galactic frame. As a massive vector boson, it encompasses three degrees of freedom: the longitudinal (L) and two transverse modes (left/right-hand circularly polarized states, $T_l$ and $T_r$, respectively). Polarization (P) basis vectors $e^\mu_P$ 
are introduced, conforming to the unitary gauge and normalization condition $e^{\mu}_{P} e_{\mu}^{P}=-1$, and takes the forms:
\begin{equation}
\begin{aligned}
    e^{\mu}_{T_l}&=(0,\lp\hat{\bm{e}}_{T_1}-\mi \hat{\bm{e}}_{T_2}\rp/\sqrt{2}),\\
   e^{\mu}_{T_r}&=(0,\lp\hat{\bm{e}}_{T_1}+\mi \hat{\bm{e}}_{T_2}\rp/\sqrt{2}),\\
    e^{\mu}_{L}&=(k/m_{X},\omega \, \hat{\bm{k}}/m_{X}).
\end{aligned}
\end{equation}
Here, $\hat{\bm{k}}$, $\hat{\bm{e}}_{T_1}$ and $\hat{\bm{e}}_{T_2}$ form a Cartesian orthogonal basis parameterized in the Galactic frame as:
\begin{equation}
\begin{aligned}
    \hat{\bm{k}} &\equiv \left(\sin \theta_k \cos \phi_k,\sin \theta_k \sin \phi_k,\cos \theta_k \right), \\
    \hat{\bm{e}}_{T_1} &\equiv \left(\cos \theta_k \cos \phi_k,\cos \theta_k \sin \phi_k,-\sin \theta_k \right), \\
    \hat{\bm{e}}_{T_2}  &\equiv \left(-\sin \phi_k,\cos \phi_k,0 \right),
    \label{eqSM:OB}
\end{aligned}
\end{equation}
using spherical coordinate $(\theta_k, \phi_k)$. The dark photon energy $\omega$ satisfies the dispersion relation $\omega^2=|\bm{k}|^2+m_{X}^2$. It's notable that the spatial components of the transverse basis, $\bm{e}_{T_{l/r}} \equiv \lp\hat{\bm{e}}_{T_1}\mp\mi \hat{\bm{e}}_{T_2}\rp/\sqrt{2}$, are unit vectors, while the longitudinal one is $|\bm{e}_L|=\omega/m_X$. Therefore, for relativistic dark photons, the longitudinal mode can produce significantly larger signals in a cavity, as will be discussed further.

A general time-dependent dark photon field is a sum of dark photons with various momenta:
\begin{equation}
    X^{ \mu} (t,\bm{x}) = \int \sum_P\, X_{P}(\omega,\hat{\bm{k}}) \,e^{\mu}_{P}\,\me^{\mi(\omega t- \bm{k}\cdot \bm{x})}\,\frac{\D \omega}{2\pi} \, \D^2 \hat{\bm{k}},
    \label{eqSM:X_freqdomain}
\end{equation}
where $X_P(\omega,\hat{\bm{k}})$ represents the corresponding field amplitude. As the production mechanisms considered in this study are based on incoherent decay of dark matter from different locations in the galaxy, the dark photon field should be a sum of incoherent waves. Thus, we treat $X_P(\omega,\hat{\bm{k}})$ as a Gaussian stochastic field, characterized by its two-point correlation function:
\begin{equation}
    \langle X_P(\omega,\hat{\bm{k}}) X_{P'}(\omega',\hat{\bm{k}}')^* \rangle = \delta(\omega-\omega') \delta^2(\hat{\bm{k}}-\hat{\bm{k}}')\delta_{PP'} \mathcal{S}_P(\omega,\hat{\bm{k}}).
    \label{eqSM:X_PSD}
\end{equation}
In this expression, $\langle \cdots \rangle$ denotes the ensemble average. $\mathcal{S}_P(\omega,\hat{\bm{k}})$ is the dark photon PSD, reflecting the dark photon spectrum from the direction $\hat{\bm{k}}$ of polarization mode $P$. The normalization of the PSD leads to the dark photon energy density 
\begin{equation}
   \rho_X = \frac{1}{4\pi^2}\int \omega^2\,\sum_P \mathcal{S}_P(\omega,\hat{\bm{k}}) \, \D \omega\, \D^2 \hat{\bm{k}}.
   \label{eqSM:rhoXS}
\end{equation}

\subsection*{B. Cavity Response to Dark Photons}
We analyze the behavior of a cavity during the transit of a dark photon, particularly in the context of kinetic mixing with electromagnetic photons $A^\mu$. The interaction Lagrangian in the interaction picture is given by $\epsilon m_X^2 A^{\mu}X_{\mu}$, where $\epsilon$ represents the kinetic mixing coefficient. In this scenario, the dark photon functions as an effective current, altering the electromagnetic field equation:
\begin{equation}
    \Box \bm{E}(t,\bm{x})=\epsilon\, m_{X}^2\partial_t\bm{X}.
    \label{eqSM:Maxwell eq}
\end{equation}
Here, the electric field $\bm{E}$ can be expressed as a sum of orthogonal eigenmodes within the cavity:
\begin{equation}
    \bm{E}(t,\bm{x})= \sum_n \xi_n(t) \bm{E}_n (\bm{x}).
    \label{eqSM:Expand}
\end{equation}
The mode basis functions $\bm{E}_n$ comply with:
\begin{equation}
    (\nabla^2+\omega_n^2) \bm{E}_n(\bm{x})=0, \qquad \int_V   \bm{E}_n \cdot  \bm{E}_m^* \, \D^3 \bm{x}=\delta_{mn},\label{eqSM:normalization}
\end{equation}
where $\omega_n$ are the resonant frequencies, and $\xi_n(t)$ describe their temporal evolutions. Integrating Eq.\,(\ref{eqSM:Expand}) and (\ref{eqSM:normalization}) into Eq.\,(\ref{eqSM:Maxwell eq}) and considering cavity dissipation, we can solve for $\xi_0$ in the frequency domain as detailed in ~\cite{SHANHE:2023kxz}:
\begin{equation}
    \xi_0(\omega) =\frac{\mi \, \epsilon \, m_{X}^2 \, \omega}{\omega^2-\omega_0^2+ \mi \omega_0 \omega /Q_L}  \sum_P \int    X_{P}(\omega,\hat{\bm{k}}) \, \D^2 \hat{\bm{k}} \int_V  \me^{-\mi \bm{k} \cdot \bm{x}} \, \bm{e}_P \cdot \bm{E}_0(\bm{x}) \,\D^3 \bm{x}.  
\end{equation}
Here, $Q_L$ is the loaded quality factor of the $\bm{E}_0$ mode. The two-point correlation function of $\xi_i(\omega)$ contributes to the signal power in the cavity as $\langle \xi_0^P(\omega) \xi_0^{P^{\prime}}(\omega^{\prime})^* \rangle =  \delta_{P P^{\prime}} \delta(\omega-\omega^{\prime})S_{\mathrm{sig}}^P(\omega)$. Utilizing Eq.\,(\ref{eqSM:X_PSD}), we obtain the expression for $S_{\mathrm{sig}}^P$:
\begin{equation}
    S_{\mathrm{sig}}^P(\omega) = \frac{\epsilon^2\, m_{X}^4\, V\, |\bm{e}_P|^2}{4(\omega-\omega_0)^2 + \omega_0^2/Q_L^2}
    \int  \mathcal{S}_P(\omega,\hat{\bm{k}}) C_{P}(\hat{\bm{k}}) \D^2 \hat{\bm{k}},
    \label{eqSM:Ssig}
\end{equation}
where we apply $Q_L \gg 1$ to simplify the Lorentz resonant factor. The dark photon overlapping form factor $C_{P}(\hat{\bm{k}})$ is defined as:
\begin{equation}
    C_{P}(\hat{\bm{k}}) \equiv \frac{1}{V} \left\vert\int_V  \me^{-\mi \bm{k}\cdot \bm{x}} \, \hat{\bm{e}}_{P}(\hat{\bm{k}})\cdot\bm{E}_0(\bm{x}\,) \,\D^3 \bm{x} \,  \right\vert^2.
    \label{eqSM:overlap}
\end{equation}
As previously discussed, the factor $|\bm{e}_P|^2$ in Eq.\,(\ref{eqSM:Ssig}) enhances the longitudinal mode's contribution to the signal for relativistic dark photons. The form factor is maximized when the dark photon wavefunction, proportional to $\hat{\bm{e}}_P$, aligns with the electric field $\bm{E}_0$. Compared to non-relativistic DPDM~\cite{SHANHE:2023kxz}, the form factor here includes an additional phase factor $\me^{-\mi \bm{k}\cdot \bm{x}}$, which could potentially suppress the signal. As the TM$_{010}$ mode of the axis-symmetric cavity in our study lacks an azimuthal component, the two transverse modes, $T_l$ and $T_r$, have identical form factors.

\section*{II: Galactic Dark Photons from Dark Matter Decay}
This section delves into the production of galactic dark photons as a result of dark matter decay. We initially explore various coupling types, each leading to distinct dominant polarization modes. When the coupling strength is significant, two-body decay can exponentially deplete the scalar dark matter background; otherwise, it may yield a dark photon flux too weak for detection. We then introduce two alternative production mechanisms, detailing how they can generate observable galactic dark photons. It should be noted that the constraints for DPDM may not apply universally to galactic dark photons. Only constraints that do not presuppose dark photons as the predominant component of dark matter should be considered relevant.

\subsection*{A. Perturbative Decay Widths}
    We examine three interaction types between a scalar field $\Phi$ and dark photons: dark Higgs-like $g_{{H}} \Phi X_\mu X^\mu$, dilaton-like $g_{{d}} \Phi X_{\mu\nu}X^{\mu \nu}$, and axion-like $g_{{a}} \Phi X_{\mu \nu}\tilde{X}^{\mu \nu}$. Assuming the scalar mass $m_\Phi$ is much larger than the dark photon mass $m_X$, the corresponding decay widths for either longitudinal or transverse modes are summarized in Table.\,\ref{tabSM:DecayWidth}. The dark Higgs-like interaction, originating from spontaneous symmetry breaking and characterized by $g_{H} = \sqrt{2}e m_X$ (with $e$ as the dark charge of the dark Higgs), predominantly decays into longitudinal modes. Both dilaton-like and axion-like interactions, frequently proposed in theories featuring extra dimensions~\cite{Damour:1994ya,Antoniadis:2001sw,Svrcek:2006yi,Arvanitaki:2009fg,Gendler:2023kjt}, mainly produce transverse modes of dark photons. In our analysis, we consider purely longitudinal or transverse modes of dark photons following dark matter decay.

\begin{table}[h]
\centering
\begin{tabular}{c|c|c|c}
     & $\quad g_{{H}} \Phi X_\mu X^\mu \quad$ & $\quad g_{{d}} \Phi X_{\mu\nu}X^{\mu \nu}\quad$  &$\quad g_a \Phi X_{\mu \nu}\tilde{X}^{\mu \nu}\quad$ \\ \hline
   $\Gamma(\Phi\rightarrow X_{T}X_{T})$  & $\frac{g_{{H}}^2}{4\pi m_\Phi}$ & $ \frac{m^4_\Phi}{2m_X^4} \frac{g_{{d}}^2 m_X^4}{2\pi m_\Phi}$  & $\frac{g_{{a}}^2 m_\Phi^3}{4\pi}$\\ \hline
   $\Gamma(\Phi\rightarrow X_{L}X_{L})$  & $ \frac{m^4_\Phi}{8m_X^4} \frac{g_{{H}}^2}{4\pi m_\Phi}$ &$\frac{g_{{d}}^2 m_X^4}{2\pi m_\Phi}$ &  /
\end{tabular}
\caption{Perturbative decay widths for three distinct interactions between a scalar particle $\Phi$ and dark photons $X^\mu$, focusing on the relativistic limit where the final-state dark photons are significantly lighter than the scalar ($m_X \ll m_\Phi$). The total decay widths are categorized based on the polarization modes of the dark photons, distinguishing between transverse and longitudinal modes.}
\label{tabSM:DecayWidth}
\end{table}

\subsection*{B. Two-body Parametric Decay}

In scenarios with a coherent oscillating scalar background $\Phi = \Phi_0 \cos m_\Phi t$, the decay rate of a scalar into dark photons with the same momentum as previously generated particles is enhanced due to Bose enhancement in the final state phase space, fostering exponential growth in dark photon density. This phenomenon is evident from the dark photon's equation of motion:
\begin{equation}
    \ddot{X} + \left( k^2+m_{X}^2+ q \cos m_\Phi t \right) X = 0,
    \label{eqSM:PREoM}
\end{equation}
where $q$ is a parameter dependent on the coupling between $\Phi$ and the dark photon. For example, $q = g_a \Phi_0 k m_\Phi$ in the case of axion-photon coupling, and $q = g_H \Phi_0$ or $g_H \Phi_0 m_\Phi^2/(2 m_X^2)$ for dark Higgs-like coupling with transverse and longitudinal modes of dark photon, respectively. The above equation (\ref{eqSM:PREoM}) is analogous to the Mathieu equation, which predicts several instability bands in momentum space~\cite{Mathieu1868,kovacic2018mathieu}. 
Within these bands, one can expect a significant exponential increase in occupation number, a process typically referred to as parametric resonance. 
The first band typically has the broadest bandwidth and can be analyzed either numerically or through semi-analytic methods like Floquet theory~\cite{stoker1950nonlinear,acar2016floquet,Hertzberg:2018zte} or non-adiabatic condition checks~\cite{Chen:2023vkq}, yielding the range:
\begin{equation}
\frac{m_\Phi^2}{4}-q \leq k^2+m_X^2 \leq \frac{m_\Phi^2}{4} + q.
\end{equation}
Under the assumption that $m_\Phi \gg m_X$, the first instability band centers around $m_\Phi$ with a bandwidth $\Delta k = g_a \Phi_0 m_\Phi$ for axion-photon coupling, and $\Delta k = 2 g_H \Phi_0/m_\Phi$ and $g_H \Phi_0 m_\Phi/m_X^2$ for the transverse and longitudinal modes in dark Higgs coupling. Both numerical and Floquet theory analyses suggest that the exponential growth rate approximately equals the bandwidth $\Delta k$. Consequently, the production of dark photons is predominantly in the longitudinal mode from dark-Higgs coupling, consistent with perturbative analysis.

Alternatively, this exponential growth can be effectively described in kinetic theory, where the decay rate is significantly amplified by the dark photon occupation number $f_X$~\cite{Tkachev:1987cd,Caputo:2018vmy,Carenza:2019vzg,Alonso-Alvarez:2019ssa,Dev:2023ijb}:
\begin{equation}
    \frac{\D n_X}{\D t} = 2n_\Phi (1+2f_X) \Gamma(\Phi \rightarrow XX) + \cdots.\label{eqSM:dndt}
\end{equation}
Here, $n_X$ and $n_\Phi$ represent the number densities of dark photons and the scalar dark matter background, respectively, with $\cdots$ denoting other terms influencing the density. In the narrow bandwidth limit, where $\Delta k \ll m_\Phi/2$, the occupation number can be approximated as $f_X \approx 2\pi^2 n_X/(k^2 \Delta k)$. The exponential growth factor, derived directly from Eq.\,(\ref{eqSM:dndt}), is $8\pi^2 n_\Phi \Gamma(\Phi \rightarrow XX)/(k^2 \Delta k)$. This factor approximates to $\Delta k$ when considering $n_\Phi \approx \rho_\Phi/m_\Phi \approx m_\Phi \Phi_0^2/2$ and $k \approx m_\Phi/2$.

We can estimate the critical energy density of dark photons, $\rho_{\text{crit}}$, at which Bose-enhanced decay becomes significant, by considering the occupation number $f_X$ to be of the order of $1$. This leads to the following expression:
\begin{equation}
    \rho_{\text{crit}} =  \int\frac{4\pi k^2 \D k}{(2\pi)^3} \sqrt{k^2+m_X^2} \approx 2.1 \times 10^{-11} \rho_\gamma.
\end{equation}
where we assume the bandwidth of the dark photon, $\Delta k$, to be $10^{-3}\,k$ and $k \approx m_\Phi/2 \approx 2\pi 1.3\,$GHz. Here, $\rho_\gamma \approx 0.26\,\mathrm{eV}/\mathrm{cm}^{3}$ represents the local electromagnetic photon density~\cite{Workman:2022ynf}. When the density of dark photons surpasses $\rho_{\text{crit}}$, their population begins to exhibit exponential growth, which could potentially disrupt the stability of the dark matter halo. To mitigate this, a suitably weak coupling constant must be chosen. For example, in the case of axion-dark photon coupling, the resulting local dark photon density from perturbative decay can be approximated as $\rho_X \sim \Gamma(\Phi \rightarrow XX)\,\rho_{\odot}^{\text{DM}}\,r_G \propto g_a^2$. In this approximation, $\rho_{\odot}^{\text{DM}} = 0.3$\,GeV/cm$^3$ is the local dark matter density, and we use $r_G = 10$\,kpc as a typical galactic scale. This suggests that $g_a$ should be less than approximately $10^{-6}$\,GeV$^{-1}$ to keep $\rho_X$ below $\rho_{\text{crit}}$. However, it's still possible for an initially small dark photon density to grow beyond $\rho_{\rm crit}$ at a later stage. To suppress such growth, one must consider dissipative effects on the dark photon density, such as an escape rate due to the finite size of the boson background or gravitational redshift from the dark matter profile~\cite{Arza:2020eik}, both of which impart an exponential decay rate with a factor of $1/R$, where $R$ denotes the typical length scale of the boson background or gravitational potential. By comparing the exponential growth rate, roughly $\Delta k$, with the decay rate of $1/R$, a similar condition is deduced for axion-photon coupling when considering gravitational redshift in the Milky Way: $g_a \ll 10^{-6}$\,GeV$^{-1}$~\cite{Arza:2020eik}. Nonetheless, within this parameter range, the energy density of the decay products of dark photons is too low to be detectable.

To address the paradox of achieving sufficiently large dark photon fluxes without depleting the dark matter halo, we consider a scenario where a fraction of the dark matter profile consists of scalar clumps. Formed due to early universe's small-scale density perturbations~\cite{Kolb:1993zz,Visinelli:2017ooc,Levkov:2018kau,Buschmann:2019icd}, these clumps are considerably denser than the average dark matter halo. The inner region of these clumps, often termed as a boson star, is surrounded by a minicluster~\cite{hogan1988axion,Kolb:1993hw},
with densities ranging between that of the boson star and the ambient halo profile. The production of dark photons through Bose-enhancement predominantly occurs within the boson star region, eventually attaining a saturation phase for both the dark photon field and the scalar background~\cite{Spieksma:2023vwl,AxionStarradiop}. The dark photon field's equilibrium is upheld by equilibrating the parametric decay rate, approximately $\Delta k$, with the escape rate from the boson star, about $1/R_{\rm BS}$, where $R_{\rm BS} \approx 1/(m_\Phi v_0)$ denotes the boson star radius:
\be \text{Saturation\ of\ dark\ photon:} \qquad  g_a \Phi_0^{\rm sat} m_\Phi = m_\Phi v_0.\label{eqSM:DPS}\ee
Here, $\Phi_0^{\rm sat}$ symbolizes the saturated boson star field value. The dark photon's bandwidth is characterized by a quality factor $Q_X = m_\Phi/(2 \Delta k) = 1/(2 v_0)$. Assuming $v_0$ approximates the typical virial velocity of the Milky Way ($\sim10^{-3}\,c$), we derive $Q_X \approx 500$. The boson star's equilibrium hinges on balancing energy acquisition from accretion in the surrounding minicluster and energy loss due to parametric decay into dark photons. The accretion rate, derived from simulations, is approximately $1/(0.3 \, \tau_{\rm rel})$, where $\tau_{\rm rel} \approx 2\sqrt{2}m_\Phi^3 v_0^2/(3\sigma_{\Phi}n_{\rm MC}^2 \pi^2)$ is the relaxation time from the minicluster~\cite{Tkachev:1991ka,Levkov:2018kau,Veltmaat:2019hou,Budker:2023sex,Jain:2023tsr}, 
characterized by number density $n_{\rm MC}$ and self-interaction cross-section $\sigma_{\Phi}^{\rm SI}$. For a quartic self-interaction, $\sigma_{\Phi} = m_\Phi^2/(8\pi f_\Phi^4)$, with $f_\Phi$ being the symmetry breaking scale or decay constant for axion-like particles. The parametric decay rate is given by $2 \, \Gamma(\Phi\rightarrow XX) \, f_X$. Consequently, the boson star's equilibrium condition is:
\be  \text{Saturation\ of\ boson\ star:} \qquad  1/(0.3\,\tau_{\rm rel}) = 2 \, \Gamma(\Phi\rightarrow XX) \, f_X^{\rm sat},\label{eqSM:BSS}\ee
where $f_X^{\rm sat}$ is the saturated dark photon occupation number. These two saturation conditions, as outlined in Eq.\,(\ref{eqSM:DPS}) and (\ref{eqSM:BSS}), can dynamically be achieved from an initially low-density boson star. The relaxation process exponentially amplifies the boson star field value, and upon reaching $\Phi_0^{\rm sat}$ as per Eq.\,(\ref{eqSM:DPS}), the dark photon starts accumulating, rapidly saturating $f_X^{\rm sat}$ and subsequently quenching the exponential growth of the boson star. During the saturation phase, the boson star continuously feeds from the minicluster, converting it into steady dark photon fluxes. To contrast with decay from a smoothly distributed dark matter halo, we introduce an effective decay width for the parametric decay process originating from the boson star:
\begin{equation}
  \Gamma_{\text{PD}} =  \frac{\eta}{0.3 \tau_{\text{rel}}},
\end{equation}
where $\eta$ represents the mass fraction of boson stars in the overall dark matter halo. The effective decay's number spectrum is presumed to be flat within the bandwidth $\Delta k = m_\Phi/(2Q_X)$:
\begin{equation}
    \frac{\D  N_{\text{PD}}}{\D \omega} = \frac{ 4 Q_X}{m_\Phi}\Theta\left(\frac{m_\Phi}{4 Q_X}-|\omega-\frac{m_\Phi}{2}|\right),
\end{equation}
where $\Theta(\cdots)$ represents the Heaviside step function.

The feasibility of the saturated boson star model can be evaluated using a set of realistic benchmark parameters. Consider, for example, a dark matter particle with mass $m_\Phi = 4\pi 1.3$\,GHz, a decay constant $f_\Phi = 10^9$\,GeV, and a minicluster number density $n_{\rm MC} = 10^{30}$\,cm$^{-3}$~\cite{Kolb:1993zz,Tkachev:2014dpa,Tinyakov:2015cgg}. Given these parameters, the relaxation time $\tau_{\rm rel}$ is calculated to be approximately $10^7$\,years, significantly shorter than the age of the universe, $t_U \approx 1.37 \times 10^{10}$\,years~\cite{Workman:2022ynf}. Setting the boson star mass fraction of the total dark matter mass to $\eta = 0.1\%$ aligns the effective decay width with the inverse of $t_U$, which is crucial for normalizing the local energy density of dark photons in the main text. If we adopt a coupling constant $g_a =10^{-11}$\,GeV$^{-1}$, the saturated boson star field value, $\Phi_0^{\rm sat}$, is predicted to be around $f_\Phi/100$, as inferred from the condition (\ref{eqSM:DPS}). This ensures the stability of the boson star, preventing its collapse due to self-interactions~\cite{Eby:2016cnq,Levkov:2016rkk}. Furthermore, this choice of $g_a$ effectively avoids triggering parametric resonance within miniclusters, assuming they typically have a radius of about $10^7$\,m~\cite{Kolb:1993zz,Tkachev:2014dpa,Tinyakov:2015cgg,Ellis:2022grh}.

\subsection*{C. Four-body Cascade Decay}
An alternative approach to circumvent the rapid depletion of the scalar dark matter background is by 
treating the scalar $\Phi$ as an intermediate state in the decay chain from dark matter $\chi$ to dark photon $X^\mu$. This transforms the decay process into a cascade sequence, represented as
\begin{equation}
\chi \rightarrow \Phi \Phi \rightarrow X^\mu X^\mu X^\mu X^\mu,
\end{equation}
which is analogous to the galactic axion production explored in Ref.\,\cite{ADMX:2023rsk}. The kinematics inherent in this four-body decay process significantly expand the final state phase space. Specifically, in the regime where $m_\Phi \ll m_{\rm DM}$ (with $m_{\rm DM}$ denoting the dark matter particle $\chi$'s mass), the spectrum of the dark photons generated in each decay instance can be approximated as a uniform distribution up to a limit of $m_{\rm DM}/2$~\cite{Elor:2015tva,Elor:2015bho}:
\begin{equation}
\frac{\mathrm{d} N_{\text{CD}}}{\mathrm{d} \omega} = \frac{8}{m_{\rm DM}} \Theta\left(\frac{m_{\rm DM}}{2} - \omega\right).
\end{equation}
The decay width of the $\chi$ is primarily governed by the initial decay process $\chi \rightarrow \Phi\Phi$, denoted as $\Gamma_{\text{CD}}$.

\subsection*{D. Angular Spectral Density}
Both the two-body parametric decay and the four-body cascade decay processes occur throughout the galaxy. To calculate the flux at Earth, it's necessary to integrate along the line of sight (l.o.s.) in a specific direction, resulting in the differential flux~\cite{Cirelli:2010xx}:
\be 
\frac{\D \Phi_X}{\D^2 \hat{\bm{k}}\, \D \omega}=\frac{r_{\odot}}{4 \pi} \frac{\rho^{\rm DM}_{\odot}}{m_{\mathrm{DM}}} J(\theta_k)  \Gamma_{\text{PD/CD}} \frac{\D N_{\text{PD/CD}}}{\D \omega}, \qquad J(\theta_k) \equiv \int_{\text {l.o.s. }} \left(\frac{\rho^{\rm DM}(r(s, \theta_k))}{\rho^{\rm DM}_{\odot}}\right) \frac{\D s}{r_{\odot}}.\label{eqSM:DF}
\ee
Note that $m_{\rm DM} = m_\Phi$ for parametric decay, while $m_{\rm DM} = m_\chi$ for cascade decay. We employ spherical coordinates in the Galactic frame, with Earth at the origin and the Galactic Center (GC) as the $-\hat{\bm{z}}$ axis, defining $\hat{\bm{k}} \equiv (\theta_k,\phi_k)$. We assume a spherically symmetric dark matter halo density $\rho^{\rm DM} (r)$ that varies with distance $r$ from the GC. The parameter $r_\odot = 8.33$\,kpc is the distance from Earth to the GC. The factor $J(\theta_k)$ quantifies the line of sight integral of the dark matter density, with $r(s, \theta_k) \equiv (r_{\odot}^2 + s^2 - 2 r_{\odot} s \cos \theta_k)^{1 / 2}$. From the differential flux (\ref{eqSM:DF}), we derive the local dark photon energy density:
\be \rho_X = \iint \frac{\D \Phi_X}{\D^2 \hat{\bm{k}}\, \D \omega} \omega\, 2\pi \sin \theta_k \, \D \theta_k\, \D \omega = \frac{1}{2}r_{\odot} \rho^{\rm DM}_{\odot} \Gamma_{\text{PD/CD}} \int {J(\theta_k)} \sin \theta_k \, \D \theta_k. \label{eqSM:rhoX} \ee
The dark matter distribution is modeled using the Navarro, Frenk, and White (NFW) profile~\cite{Navarro:1995iw}:
\be \rho^{\rm DM} (r) = \rho_s \frac{r_s}{r} (1+\frac{r}{r_s})^{-2},\label{eqSM:rhoDM}\ee
with scale parameters $r_s = 24.42$\,kpc and $\rho_s = 0.184$\,GeV/cm$^3$. For the scenario involving parametric decay from boson stars, it is assumed that the distribution of the boson stars adheres to the NFW profile. We adopt the assumption that the effective decay rate $\Gamma_{\text{PD/CD}}$ attains its maximum value consistent with existing cosmological constraints~\cite{Workman:2022ynf,DES:2020mpv}:
\be \Gamma_{\text{PD/CD}}^{-1} \approx 3.6\,t_U \approx 4.97\times 10^{10}\,\text{years}.\label{eqSM:GammaAU}\ee
Incorporating Eq.\,(\ref{eqSM:rhoDM}) and (\ref{eqSM:GammaAU}) into Eq.\,(\ref{eqSM:rhoX}) yields a maximally allowed flux of dark photons, approximately $\rho_X \approx 1000\,\rho_\gamma$~\cite{ADMX:2023rsk}.

Comparing Eq.\,(\ref{eqSM:rhoXS}) and (\ref{eqSM:rhoX}), we derive the relationship between the angular spectral density and differential flux:
\be \mathcal{S}_P(\omega, \hat{\bm{k}}) = \frac{4\pi^2}{\omega}\,\frac{\D \Phi_X}{\D^2 \hat{\bm{k}}\, \D \omega}. \ee
This relationship forms the basis for two benchmark angular spectral densities discussed in the maintext:
\begin{equation}
\begin{aligned}
    \mathcal{S}_P^{\text{PD}}(\omega, \hat{\bm{k}}) =&\, \frac{4\pi\, Q\, \rho_X}{\omega\, m_{\rm DM}^2\, \bar{J}} \,J(\theta_{k})\,\Theta\left(\frac{m_{\rm DM}}{4 Q}-|\omega-\frac{m_{\rm DM}}{2}|\right),\\
    \mathcal{S}_P^{\text{CD}}(\omega, \hat{\bm{k}}) =&\, \frac{8\pi\, \rho_X}{\omega \,m_{\rm DM}^2\,\bar{J}} \, J(\theta_{k})\,\Theta\left(\frac{m_{\rm DM}}{2} - \omega\right),
    \end{aligned}
\end{equation}
where $\bar{J} \equiv \int J(\hat{\bm{k}})\D^2 \hat{\bm{k}}/(4\pi) \approx 2.19$.

To calculate the modulation of galactic dark photon signals, it is necessary to consider the variation of the cavity's symmetry axis in the Galactic frame, denoted as $\hat{\bm{l}}_{\rm cav}(t) \equiv (\theta_{\rm cav}(t), \phi_{\rm cav}(t))$. For simplicity, we set the azimuthal angle of the Earth's rotation axis at $0$, leading to the following geometric relationships:
\begin{equation}
\begin{aligned}
    \cos \theta_{\rm cav}(t)=& \cos \, \delta_{\rm GC} \, \cos \, \varphi_{\rm cav} \, \cos \left( 2 \pi t/24 + \lambda_{\rm cav}-\alpha_{\rm GC}\right) + \sin \, \delta_{\rm GC} \, \sin \, \varphi_{\rm cav},\\
    \sin \theta_{\rm cav}(t) \sin \phi_{\rm cav}(t)=& \cos \varphi_{\rm cav} \sin \left( 2 \pi t/24 + \lambda_{\rm cav}-\alpha_{\rm GC}\right).
        \end{aligned}
\end{equation}
In this context, $(\varphi_{\rm cav}, \lambda_{\rm cav}) = (39.9^{\circ}, 116.4^{\circ})$ represent the latitude and longitude of the cavity, which is oriented towards the Earth's center throughout the experiment. The coordinates $(\delta_{\rm GC}, \alpha_{\rm GC}) = (-29.0^{\circ}, 266.4^{\circ})$ specify the declination and right ascension of the GC in the celestial sphere. The time variable $t$ is expressed in hours, based on the J2000 standard time system~\cite{SOFA:2021-01-25}. Practically, due to the axial symmetry of the dark photon angular spectrum, the modulation of the cavity signal does not exhibit sensitivity to variations in $\phi_{\rm cav}(t)$.

\section*{III: Experimental Operations}

This section outlines the experimental procedures and the calibration of relevant parameters.

We conducted two continuous operational periods from April 12-14 and April 23-26, 2023, executing $N_{\rm bin}=350$ and $800$ scan steps respectively. In each step, the I/Q mode of the spectrum analyzer recorded the complex voltage signals from the readout antenna, encompassing a data recording phase and dual calibration phases for the resonant frequency.

During each scan step labeled by $i$, a broadband noise source was initially used to calibrate the resonant frequency $\omega_0^i$ by identifying the peak in the power spectral density (PSD). Subsequently, the noise source was deactivated, and the time-domain voltage signal was recorded over an integration time of $t_{\rm int}=100\,\mathrm{s}$, converting into the received power $\mathcal{P}^i$. We recorded the end time $t^i$ of each integration phase via a computer linked to the spectrum analyzer. Assuming minimal variation in $\hat{\bm{l}}_{\rm cav}(t)$ within the short integration period, we evaluated the expected signal $\mathcal{P}^i_{\rm sig}$ at $t^i$. Post-recording, we recalibrated the frequency to mitigate any shifts smaller than $\Delta \omega_0$ and adjusted the cavity using a step motor-controlled tuning arm~\cite{Pischalnikov:2015eye,Pischalnikov:2019iyu} to change the resonant frequency by roughly $1.3\,\mathrm{kHz}$ before proceeding.

The calibration parameters are detailed in Table.\,\ref{tabSM:parameters}. Daily, $G_{\rm net}$, $\beta$, and $Q_L$ were calibrated using a vector network analyzer (VNA) and a vertical test stand (VTS) system~\cite{Melnychuk:2014pka}. Stability in measured values allowed the usage of average values in analysis, accounting for standard deviation as uncertainty. The cavity volume $V$ and the angular-dependent overlap factor $C_P(\hat{\bm{k}})$ were determined via COMSOL simulations. We conservatively estimate the fractional uncertainties in $V$ at $0.5\%$, arising from inner cavity wall displacements caused by acid pickling and frequency-tuning deformations. Additionally, the potential deviation of the cavity’s $z$-axis contributes up to $0.88\%$, as detailed in the next section, resulting in a total uncertainty of $C_P$ of $1.0\%$.


We determined the analysis bandwidth based on two phenomena affecting the resonant frequency: frequency drift, a gradual variation peaking at about $ 1.5$\,Hz during $t_{\rm int}$ due to mechanical resistance, and microphonics, oscillating deviations with an RMS of $4.1$\,Hz~\cite{Pischalnikov:2019iyu}. Combining these, we set an analysis bandwidth of $\Delta \omega_0/(2\pi)=  11.5 \,\mathrm{Hz}$ to maintain the resonant band within this limit. Detailed values and uncertainties for all parameters are provided in Table.\,\ref{tabSM:parameters}.

\begin{table}[htb]
\centering
\begin{tabular}{l c c}
\hline
\hline
        & Value                 & Fractional uncertainty $F_j$  \\
\hline
$V$              & $3900$\,mL                 & $< 0.5\%$   \\
$C_P(\hat{\bm{k}})$              & $0 \sim 0.28 $                 & $1.0\%$  \\
$\beta$          & $0.634\pm 0.014$                  & $1.4\%$  \\
$G_{\rm net}$              & $(57.30 \pm 0.14)$\,dB               & $3.1\%$ \\
$Q_L$            & $(9.092\pm 0.081)\times 10^9$  & $\slash$  \\
$t_{\rm int}$            & $100\,\mathrm{s}$  & $\slash$  \\
$\Delta \omega_0/(2\pi)$            & $11.5 \,\mathrm{Hz}$  & $\slash$  \\
\hline
\hline
\end{tabular}
\caption{Calibrated parameters for the SRF cavity and amplifier, showcasing average values, uncertainties, and fractional uncertainties $F_j$ impacting the dark photon-induced signal in this study.}
\label{tabSM:parameters}
\end{table}

\section*{IV: Data Analysis}

The data for this study are presented in Fig.\,2 in the main text. Given the Gaussian nature of the measured data, we utilize the following probability function:
\begin{equation}
    \mathrm{Pr}\left(\delta^i \mid \epsilon, m_{X}\right) \propto \prod_{i} \exp \left( \frac{-\left( \delta^i/\tilde{\delta}_{\rm sig}^i -\left( \epsilon/\epsilon_{\rm ref} \right)^2 \right)^2}{2 (\sigma_{\rm tot}^i)^2} \right),
    \label{eqSM:likelihood}
\end{equation}
where $\tilde{\delta}_{\rm sig}^i \equiv \delta_{\rm sig}^i|_{\epsilon=\epsilon_{\rm ref}}$ is introduced alongside an arbitrary reference value $\epsilon_{\rm ref}$ to account for both measurement error and uncertainties from calibrated parameters. Employing the error propagation formula, we deduce the combined uncertainty for $\delta^i/\tilde{\delta}_{\rm sig}^i$, labeled as $\sigma_{\rm tot}^i$:
\begin{equation}
    \left(\sigma_{\rm tot}^i\right)^2 =  \frac{1}{\left(\tilde{\delta}_{\rm sig}^i \right)^2} + \frac{(\delta^i)^2}{\left(\tilde{\delta}_{\rm sig}^i \right)^4} \sum_j \left(\frac{\partial \tilde{\delta}_{\rm sig}^i}{\partial j} \sigma_j \right)^2.
    \label{eqSM:error_prop}
\end{equation}
Here, $j$ represents various calibrated parameters with $\sigma_j$ as their respective uncertainties. The first term in Eq.\,(\ref{eqSM:error_prop}) signifies the measurement error, assuming a standard deviation of $1$ for $\delta^i$.
Substituting Eq.\,(\ref{eqSM:error_prop}) into Eq.\,(\ref{eqSM:likelihood}) simplifies the probability function to:
\begin{equation}
    \mathrm{Pr} \left(\delta^i \mid \epsilon, m_{X}\right) \propto \prod_{i} \exp \left( \frac{-\left( \delta^i -\delta_{\rm sig}^i \right)^2}{2 + 2(\delta^i)^2 \sum_j F_j^2} \right),
    \label{eqSM:PF}
\end{equation}
as utilized in the main text. The fractional uncertainty is defined as:
\begin{equation}
    F_j \equiv \frac{\partial \tilde{\delta}_{\rm sig}^i}{\partial j} \frac{\sigma_j}{\tilde{\delta}_{\rm sig}^i}.
    \label{eqSM:Fj}
\end{equation}
The relevant contributions include $j=G_{\rm net}$, $\beta$, $V$, and $C_P$:
\begin{equation}
    \sum_j F_j^2 =  \frac{\sigma_V^2}{V^2}+\frac{\sigma_{C_P}^2}{C_P^2}+\left(\frac{1}{\beta}-\frac{1}{1+\beta}\right)^2\sigma_{\beta}^2+\frac{\sigma_{G_{\text{net}}}^2}{G_{\text{net}}^2}
\end{equation}
with mean values and uncertainties presented in Table.\,\ref{tabSM:parameters}. Note that $G_{\rm net}$ is evaluated as $(57.30 \pm 0.14)$\,dB$\rightarrow 10^{(5.730 \pm 0.014)}$.

Using the probability function (\ref{eqSM:PF}), we can determine the 90\% exclusion limit for the kinetic mixing parameter, denoted as $\epsilon_{90\%}$, by solving:
\begin{equation}
    \frac{\int_0^{\epsilon_{90\%}^2} \mathrm{Pr}\left(\delta^i \mid \epsilon, m_{X}\right) \D \epsilon^2}{\int_0^{\infty} \mathrm{Pr}\left(\delta^i \mid \epsilon, m_{X}\right) \D \epsilon^2} = 90\%.
\end{equation}

In addition to the detailed analysis provided, we also estimate the sensitivity using the Signal-to-Noise Ratio (SNR) to demonstrate our detection capability. We employ the Dicke radiometer equation~\cite{Dicke:1946glx} for the normalized power excess:
\begin{equation}
   {\rm SNR}^2 = \sum_i {\rm SNR}^2(t^i) = \frac{t_{\rm int}\Delta \omega_0}{2\pi }\sum_i(\delta_{\rm sig}^i)^2.
\end{equation}
We introduce a modulation factor, $\mathcal{F}_{C_P}$, to assess the power excess over the average:
\begin{equation}
    \mathcal{F}_{C_P} = \frac{\sqrt{\sum_i[S_{\mathrm{sig}}^P(\omega_0,t^i)-\bar{S}_{\mathrm{sig}}^P(\omega_0)]^2/N_{\rm bin}}}{\bar{S}_{\mathrm{sig}}^P(\omega_0)}.
\end{equation}
Here, $\bar{S}_{\mathrm{sig}}^P(\omega_0)$ represent the mean signal values, accounting for the angular distribution of the dark photon source and Earth’s rotation.
In our experiment, the cavity’s $z$-axis at time $t_i$, denoted as $\hat{\bm{l}}_{\rm cav}^i$, can deviate from the vertical direction by up to $3^{\circ}$, based on engineering resolution. To assess the impact of this deviation on the form factor $C_P$, we varied $\hat{\bm{l}}_{\rm cav}^i$ within a $3^\circ$ cone. This resulted in a maximum deviation of the modulation factor $\mathcal{F}_{C_P}$ of $0.88\%$ for the longitudinal mode and $0.59\%$ for the transverse mode. Combined with the volume uncertainty of approximately $0.5\%$, the fractional uncertainty of $C_P(\hat{\bm{k}})$ is conservatively estimated as $\sqrt{(0.5\%)^2+(0.88\%)^2} \approx 1.0\%$.

Notably, the average daily overlap factor, $\bar{C}_P$, is $0.135$ ($0.066$) for the transverse (longitudinal) mode of dark photon flux from dark matter decay. From these values, the SNR for $N_{\rm bin}$ independent tests distributed evenly over the experimental period is given by:
\begin{equation}
     {\rm SNR}^2 =  \frac{\pi^3\epsilon^4 m_X^8 \rho_X^2 Q_X^2 V^2|\bm{e}_P|^4 {N_{\rm bin}} t_{\rm int}\bar{C}_P^2\mathcal{F}_{C_P}^2}{8\omega_0^{6}T_{N}^2\Delta \omega_0}.
\end{equation}
The effective noise temperature, $T_N$, is related to the noise power, $\sigma_{\mathcal{P}}$, via $\sigma_{\mathcal{P}} = T_{N}\Delta\omega_0/(2\pi)$. For simplicity, $\eta_{\rm bin}$ and $\beta$ are both approximated as $1$.

\end{document}